\documentclass{article}
\usepackage{fortschritte}

\def\sk{\vspace{.4cm}}
\def\a{{\hat a}}
\def\b{{\hat b}}
\def\c{{\hat c}}

\def\eq{\begin{equation}}
\def\en{\end{equation}}
\def\L{{\cal L}}
\def\al{{\alpha}}

\def\vphi{{\varphi}}
\def\ddelta{\delta}
\def\noi{\noindent}
\begin{document}
\def\titleline{Unified Theories
 from Fuzzy Extra Dimensions}

\def\email_speaker{George.Zoupanos@cern.ch}


\def\authors{
P.~Aschieri\1ad\2ad\3ad,  J.~Madore\4ad,
P.~Manousselis\5ad and G.~Zoupanos\2ad\5ad\6ad\sp}

\def\addresses{
\1ad
Dipartimento di Scienze e Tecnologie
Avanzate,\\ Universit\'a del Piemonte Orientale, and INFN,\\ Corso
Borsalino 54, I-15100,  Alessandria, Italy\\
\2ad
Max-Planck-Institut f\"{u}r Physik\\ F\"{o}hringer Ring 6,
D-80805 M\"{u}nchen\\
\3ad
Sektion Physik,
Universit\"{a}t M\"{u}nchen\\ Theresienstra\ss e 37, D-80333
M\"{u}nchen\\
\4ad
Laboratoire de Physique
Th\'{e}orique\\Universit\'{e} de Paris-Sud, B\^{a}timent 211,
F-91405 Orsay\\
\5ad
Physics Department National
Technical University\\ Zografou Campus, GR-15780 Athens\\
\6ad
 Theory Division, CERN, CH-1211, Geneva 23,
Switzerland
}

\def\abstracttext{
We combine and exploit ideas from Coset Space Dimensional
Reduction (CSDR) methods and
Non-commutative Geometry. We consider the dimensional reduction of
gauge theories defined in high dimensions where the compact directions are
a fuzzy space (matrix manifold). In the  CSDR one assumes that
the form of space-time is $M^{D}=M^{4} \times S/R$ with $S/R$ a
homogeneous space. Then a gauge theory with gauge group $G$
defined on $M^{D}$ can be dimensionally reduced
to $M^{4}$ in an
elegant way using the symmetries of $S/R$, in particular
the resulting four dimensional gauge is a subgroup of $G$.
In the present work we
show that one can apply the CSDR ideas in the case where the
compact part of the space-time is a finite approximation of the
homogeneous space $S/R$, i.e. a fuzzy coset. In particular
we study the fuzzy sphere case.}


\large
\makefront

\section{Introduction}
Coset Space Dimensional Reduction (CSDR) \cite{Forgacs:1979zs,
Kapetanakis:hf} is a unification scheme for obtaining realistic
particle models from gauge theories on higher $D$-dimensional
spaces $M^D$. It suggests that a unification of the gauge and
Higgs sectors of the Standard Model can be achieved in higher than
four dimensions. Moreover the addition of fermions in the
higher-dimensional gauge theory leads naturally, after CSDR, to
Yukawa couplings in four dimensions.

We study  CSDR in the non-commutative context and set the rules
for constructing new particle models that might be
phenomenologically relevant. One could study CSDR with the whole
parent space $M^D$ being non-commutative or with just
non-commutative Minkowski space or non-commutative internal space.
We specialize here to this last situation and therefore eventually
we obtain Lorentz covariant theories on commutative Minkowski
space. We further specialize to fuzzy non-commutativity, i.e. to
matrix type non-commutativity. Thus, following \cite{us}, we
consider non-commutative spaces like those studied in refs.
\cite{DV.M.K., M., Madore} and implementing the CSDR principle on
these spaces we obtain new
particle models.

\section{Fuzzy sphere}
The fuzzy sphere \cite{Mad, Madore} is a matrix approximation of
the usual sphere $S^2$. The algebra of functions on $S^2$ (for
example spanned by the spherical harmonics) is truncated at a
given frequency and thus becomes finite dimensional. The
truncation has to be consistent with the associativity of the
algebra and this can be nicely achieved relaxing the commutativity
property of the algebra. The fuzzy sphere is the ``space''
described by this non-commutative algebra. The algebra itself is
that of $N\times N$ matrices. More precisely, the algebra of
functions on the ordinary sphere can be generated by the
coordinates of {\bf{R}}$^3$ modulo the relation $
\sum_{\hat{a}=1}^{3} {x}_{\hat{a}}{x}_{\hat{a}} =r^{2}$. The fuzzy
sphere $S^2_{F}$ at fuzziness level $N-1$ is the non-commutative
manifold whose coordinate functions $i {X}_{\hat{a}}$ are $N
\times N$ hermitian matrices proportional to the generators of the
$N$-dimensional representation of $SU(2)$. They satisfy the
condition $ \sum_{\hat{a}=1}^{3} {X}_{\hat{a}} {X}_{\hat{a}} = \al
r^{2}$ and the commutation relations
\begin{equation}
[ X_{\hat{a}}, X_{\hat{b}} ] = C_{\hat{a} \hat{b} \hat{c}}
X_{\hat{c}}~,
\end{equation}
where $C_{\a\b\c}=\varepsilon_{\a\b\c}/r$ while
the proportionality factor $\al$ goes as $N^2$ for $N$ large.
Indeed it can be proven that for $N\rightarrow \infty$
one obtains the usual commutative sphere.

On the fuzzy sphere there is a natural $SU(2)$ covariant
differential calculus. This calculus is three-dimensional and the
derivations $e_\a$ along $X_\a$ of a function $ f$ are given by
$
e_\a({f})=[X_{\hat{a}}, {f}]\,.\label{derivations}
$
Accordingly the action of the Lie derivatives on functions is
given by
\begin{equation}\label{LDA}
{\cal L}_{\hat{a}} f = [{X}_{\hat{a}},f ]~;
\end{equation}
these Lie derivatives satisfy the Leibniz rule and the $SU(2)$ Lie algebra relation
\begin{equation}\label{LDCR}
[ {\cal L}_{\hat{a}}, {\cal L}_{\hat{b}} ] = C_{\hat{a} \hat{b}
\hat{c}} {\cal L}_{\hat{c}}.
\end{equation}
In the $N \rightarrow \infty$ limit the derivations $e_\a$ become
$
e_{\hat{a}} = C_{\hat{a} \hat{b} \hat{c}} x^{\hat{b}}
\partial^{\hat{c}}\,
$
and only in this commutative limit the tangent space becomes two
dimensional. The exterior derivative is given by
\begin{equation}
d{f} = [X_\a,{f}]\theta^{\hat{a}}
\end{equation}
with $\theta^{\hat{a}}$ the one-forms dual to the vector fields
$e_{\hat{a}}$, $<e_\a,\theta^\b>=\delta_\a^\b$. The space of
one-forms is generated by the $\theta^\a$'s in the sense that for
any one-form $\omega=\sum_i f_i (d h_i)\,t_i$ we can always write
$\omega=\sum_{\a=1}^3{\omega}_\a\theta^\a$ with given functions
${\omega}_\a$ depending on the functions ${f}_i$,  ${h}_i$ and
${t}_i$.
The action of
the Lie derivatives on one-forms is given by
\begin{equation}\label{2.16}
{\cal L}_{\hat{a}}(\theta^{\hat{b}}) =  C_{\a\b\c}
\theta^{\hat{c}}
\end{equation}
and it is easily seen to commute with the
exterior differential $d$. On a general one-form
$\omega=\omega_{\hat{a}}\theta^{\hat{a}}$ we have
$
{\cal L}_{\hat{b}}\omega={\cal
L}_{\hat{b}}(\omega_{\hat{a}}\theta^{\hat{a}})=
\left[X_{\hat{b}},\omega_{\hat{a}}\right]\theta^{\hat{a}}-\omega_{\hat{a}}C^{\hat{a}}_{\
\hat{b} \hat{c}}\theta^{\hat{c}}
$
and therefore
\begin{equation}
({\cal
L}_{\hat{b}}\omega)_{\hat{a}}=\left[X_{\hat{b}},\omega_{\hat{a}}\right]-
\omega_{\hat{c}}C^{\hat{c}}_{\ \hat{b}  \hat{a}}~;\label{fund}
\end{equation}
this formula will be fundamental for formulating the CSDR
principle on fuzzy cosets.

The differential geometry on  the product space Minkowski times
fuzzy sphere, $M^{4} \times S^2_{F}$, is easily obtained from that
on $M^4$ and on $S^2_F$. For example a one-form $A$ defined on
$M^{4} \times S^2_{F}$ is written as
\begin{equation}\label{oneform}
A= A_{\mu} dx^{\mu} + A_{\hat{a}} \theta^{\hat{a}}
\end{equation}
with $A_{\mu} =A_{\mu}(x^{\mu}, X_{\hat{a}} )$ and $A_{\a}
=A_{\a}(x^{\mu}, X_{\hat{a}} )$.

One can also introduce spinors on the fuzzy sphere and study the
Lie derivative on these spinors. Although here we have sketched
the differential geometry on the fuzzy sphere,  one can study
other (higher dimensional) fuzzy spaces (e.g. fuzzy $CP^M$) and
with similar techniques their differential geometry.
\section{CSDR over fuzzy coset spaces}
First we consider
 on  $M^{4} \times (S/R)_{F}$ a non-commutative
gauge theory with
gauge group $G=U(P)$
and examine its four dimensional interpretation.
The action is
\begin{equation}
{\cal A}_{YM}={1\over 4} \int d^{4}x\, Tr\, tr_{G}\, F_{MN}F^{MN},
\end{equation}
where $Tr$ denotes integration over the fuzzy coset
$(S/R)_F\,$ described by $N\times N$ matrices,
and $tr_G$ is the gauge group
$G$ trace. The higher-dimensional field strength $F_{MN}$
decomposed in four-dimensional space-time and extra-dimensional
components reads as follows
$(F_{\mu \nu}, F_{\mu \hat{b}}, F_{\hat{a} \hat{b} })\,;$
\def\phi{\vphi}
\noindent explicitly the various components of the field strength
are given by
\begin{eqnarray}
F_{\mu \nu} &=&
\partial_{\mu}A_{\nu} -
\partial_{\nu}A_{\mu} + [A_{\mu}, A_{\nu}],\\[.3 em]
F_{\mu \hat{a}} &=&
\partial_{\mu}A_{\hat{a}} - [X_{\hat{a}}, A_{\mu}] + [A_{\mu},
A_{\hat{a}}], \nonumber\\[.3 em]
F_{\hat{a} \hat{b}} &=&   [ X_{\hat{a}}, A_{\hat{b}}] - [
X_{\hat{b}}, A_{\hat{a}} ] + [A_{\hat{a}} , A_{\hat{b}} ] -
C^{\hat{c}}_{\ \hat{a} \hat{b}}A_{\hat{c}}.
\end{eqnarray}

\noi Under an infinitesimal $G$ gauge transformation
$\lambda=\lambda(x^\mu,X^\a)$ we have
\begin{equation}
\ddelta A_{\hat{a}} = -[ X_{\hat{a}}, \lambda] +
[\lambda,A_{\hat{a}}]~,
\end{equation}
thus $F_{MN}$ is covariant under {local} $G$ gauge
transformations: $F_{MN}\rightarrow F_{MN}+[\lambda, F_{MN}]$.
This is an infinitesimal abelian $U(1)$ gauge transformation if
$\lambda$ is just an antihermitian function of the coordinates
$x^\mu, X^\a$ while it is an infinitesimal nonabelian $U(P)$ gauge
transformation if $\lambda$ is valued in ${\rm{Lie}}(U(P))$, the
Lie algebra of hermitian $P\times P$ matrices. In the following we
will always assume ${\rm{Lie}}(U(P))$ elements to commute with the
coordinates $X^\a$. In fuzzy/non-commutative gauge theory and in
Fuzzy-CSDR a fundamental role is played by the covariant
coordinate,
\begin{equation}
\vphi_{\hat{a}} \equiv X_{\hat{a}} + A_{\hat{a}}~.
\end{equation}
This field transforms indeed covariantly under a gauge
transformation,
$
\ddelta(\vphi_{\hat{a}})=[\lambda,\vphi_{\hat{a}}]~.
$
In terms of $\vphi$ the field strength in the non-commutative
directions reads,
\begin{eqnarray}
F_{\mu \hat{a}} &=&
\partial_{\mu}\vphi_{\hat{a}} + [A_{\mu}, \vphi_{\hat{a}}]=
D_{\mu}\vphi_{\hat{a}},\\[.3 em]
F_{\hat{a} \hat{b}} &=&
[\phi_{\hat{a}}, \phi_{\hat{b}}] - C^{\hat{c}}_{\ \hat{a} \hat{b}}
\phi_{\hat{c}}~;
\end{eqnarray}
and using these expressions the action reads
\begin{equation}
{\cal A}_{YM}= \int d^{4}x\, Tr\, tr_{G}\,\left( {1\over 4}F_{\mu
\nu}^{2} + {1\over 2}(D_{\mu}\phi_{\hat{a}})^{2} -
V(\phi)\right),\label{theYMaction}
\end{equation}
where the potential term $V(\phi)$ is the $F_{\hat{a} \hat{b}}$
kinetic term (recall $F_{\a\b}$ is antihermitian so that $V(\phi)$
is hermitian and non-negative)
$$V(\phi)=-{1\over 4} Tr\,tr_G \sum_{\hat{a} \hat{b}} F_{\hat{a}
\hat{b}} F_{\hat{a} \hat{b}}
%
%
=-{1\over 4} Tr\,tr_G \left( [\phi_{\hat{a}},
\phi_{\hat{b}}][\phi^{\hat{a}}, \phi^{\hat{b}}] - 4C_{\hat{a}
\hat{b} \hat{c}} \phi^{\hat{a}} \phi^{\hat{b}} \phi^{\hat{c}} +
2r^{-2}\phi^{2} \right).
$$
This action is naturally interpreted as an action in four
dimensions. The infinitesimal $G$ gauge transformation with gauge
parameter $\lambda(x^\mu,X^\a)$ can indeed be interpreted just as
an $M^4$ gauge transformation. We write \eq
\lambda(x^\mu,X^\a)=\lambda^\alpha(x^\mu,X^\a){\cal T}^\alpha
=\lambda^{h, \al}(x^\mu)T^h{\cal T}^\alpha~,
\en
where ${\cal T}^\alpha$ are hermitian generators of $U(P)$,
$\lambda^\alpha(x^\mu,X^\a)$ are $N\times N$ antihermitian
matrices and thus are expressible as $\lambda(x^\mu)^{\al , h}T^h$
with $T^h$  antihermitian generators of $U(N)$. Now the Lie
algebra is  the tensor product of  ${\rm{Lie}}(U(N))$ and
${\rm{Lie}}(U(P))$, it is indeed ${\rm{Lie}}(U(N_{}P))\,$.
Similarly we rewrite the gauge field $A_\nu$ as
$
A_\nu(x^\mu,X^\a)=A_\nu^\alpha(x^\mu,X^\a){\cal T}^\alpha
=A_\nu^{h, \al}(x^\mu)T^h{\cal T}^\alpha~
$
and interpret it as a ${\rm{Lie}}(U(N_{}P))$ valued gauge field
on $M^4$. Similarly we treat $\vphi_\a$.

Up to now we have just
performed a fuzzy Kaluza-Klein reduction. Indeed in the
commutative case the expression (\ref{theYMaction}) corresponds to rewriting the initial lagrangian on $M^4\times S^2$ using
spherical harmonics on $S^2$. Here the space of functions is
finite dimensional and therefore the infinite tower of modes
reduces to the finite sum given by $Tr$.

\sk

Next we  reduce the
number of gauge fields and scalars in the action
(\ref{theYMaction}) by applying the CSDR scheme.
Since Lie $SU(2)$
acts on the fuzzy sphere $(SU(2)/U(1))_F$, and more in general  the
group $S$ acts on the fuzzy coset $(S/R)_F$,
we can state the CSDR
principle in the same way as in the continuum case, i.e. the
fields in the theory must be invariant under the infinitesimal
$SU(2)$, respectively $S$, action up to an infinitesimal gauge transformation
\begin{equation}
{\cal L}_{\hat{b}} \phi =\delta^{W_\b}\phi= W_{\hat{b}} \phi,
~~~~~~~~~~~ {\cal L}_{\hat{b}}A = \delta^{W_\b}A=-DW_{\hat{b}},
\label{csdr}
\end{equation}
where $A$ is the one-form gauge potential $A = A_{\mu}dx^{\mu} +
A_{\hat{a}} \theta^{\hat{a}}$, and $W_\b$ depends only on the
coset coordinates $X^\a$ and (like $A_\mu, A_a$) is antihermitian.
We thus write $W_\b=W_\b^\alpha{\cal T}^\alpha, \,\alpha=1,2\ldots
P^2,$ where ${\cal  T}^i$ are hermitian generators of $U(P)$ and
$(W_b^i)^\dagger=-W_b^i$, here ${}^\dagger$ is hermitian
conjugation on the $X^\a$'s. Now in order to solve the constraints
(\ref{csdr}) we cannot follow the strategy adopted in the
commutative case, where the constraints were studied just at one
point of the coset (say $y^a=0$). This is due to the intrinsic
nonlocality of the constraints. On the other hand the specific
properties of the fuzzy case (e.g. the fact that partial
derivatives are realized via commutators, the concept of covariant
coordinate) allow to simplify and eventually solve the
constraints. Indeed in terms of the covariant coordinate
$\vphi_{\hat{d}} =X_{\hat{d}} + A_{\hat{d}}$ and of
\begin{equation}
\omega_{\hat{a}} \equiv X_{\hat{a}} - W_{\hat{a}}~,
\end{equation}
the CSDR constraints assume a particularly simple form, namely
\begin{equation}\label{3.19}
[\omega_{\hat{b}}, A_{\mu}] =0,
\end{equation}
\begin{equation}\label{eq7}
C_{\hat{b} \hat{d} \hat{e}} \vphi^{\hat{e}} = [\omega_{\hat{b}},
\vphi_{\hat{d}} ].
\end{equation}
In addition we  have a consistency condition  following from the
relation $[\L_a,\L_\b]=C_{\a\b}^{~~\c}\L_\c$:
\begin{equation}\label{3.17}
[ \omega_{\hat{a}} , \omega_{\hat{b}}] = C_{\hat{a} \hat{b}}^{\ \
\hat{c}} \omega_{c},
\end{equation}
where $\omega_{\hat{a}}$ transforms as
$
\omega_\a\rightarrow \omega'_{\hat{a}} = g\omega_{\hat{a}}g^{-1}.
$
One proceeds in a similar way for the spinor fields \cite{us}.
\section{Solving the  CSDR constraints for the fuzzy sphere}
\noindent We consider  $(S/R)_{F}=S^2_{F}$, i.e. the fuzzy sphere,
and to be definite at fuzziness level $N-1$ ($N \times N$
matrices). We study first the basic example where the gauge group
$G=U(1)$. In this case the
$\omega_{\hat{a}}=\omega_{\hat{a}}(X^\b)$  appearing in the
consistency condition (\ref{3.17}) are $N \times N$
antihermitian matrices and therefore can be interpreted as
elements of ${\rm{Lie}}(U(N))$. On the other hand the
$\omega_{\hat{a}}$ satisfy the commutation relations (\ref{3.17})
of ${\rm{Lie}}(SU(2))$. Therefore in order to satisfy the
consistency condition (\ref{3.17}) we have to embed
${\rm{Lie}}(SU(2))$ in ${\rm{Lie}}(U(N))$. Let $T^h$ with $h =
1, \ldots ,(N)^{2}$ be the generators of ${\rm{Lie}}(U(N))$ in
the fundamental representation, we can always use the convention
$h= (\hat{a} , u)$ with $\hat{a} = 1,2,3$ and $u= 4,5,\ldots,
N^{2}$ where the $T^\a$ satisfy the $SU(2)$ Lie algebra,
\begin{equation}
[T^{\hat{a}}, T^{\hat{b}}] = C^{\hat{a} \hat{b}}_{\ \
\hat{c}}T^{\hat{c}}~.
\end{equation}
Then we define an embedding by identifying \eq
 \omega_{\hat{a}}= T_{\hat{a}}.
\label{embedding}
\en
The constraint (\ref{3.19}), $[\omega_{\hat{b}} , A_{\mu}] = 0$,
then implies that the four-dimensional gauge group $K$ is the
centralizer of the image of $SU(2)$ in $U(N)$, i.e. $$
K=C_{U(N)}(SU((2))) = SU(N-2) \times U(1)\times U(1)~, $$  where
the last $U(1)$ is the $U(1)$ of $U(N)\simeq SU(N)\times
U(1)$. The functions $A_{\mu}(x,X)$ are arbitrary functions of $x$
but the $X$ dependence is such that $A_{\mu}(x,X)$ is
${\rm{Lie}}(K)$ valued instead of ${\rm{Lie}}(U(N))$, i.e.
eventually we have a four-dimensional gauge potential $A_\mu(x)$
with values in ${\rm{Lie}}(K)$. Concerning the constraint
(\ref{eq7}), it is satisfied by choosing \eq \label{soleasy}
\vphi_\a=r \vphi(x) \omega_\a~,
\en
i.e. the unconstrained degrees of freedom correspond to the scalar
field $\vphi(x)$ which is a singlet under the four-dimensional
gauge group $K$.


The choice (\ref{embedding}) defines one of the possible embedding
of ${\rm{Lie}}(SU(2))$ in ${\rm{Lie}}(U(N))$. For example we
could also embed ${\rm{Lie}}(SU(2))$ in ${\rm{Lie}}(U(N))$
using
the irreducible $N$ dimensional rep. of $SU(2)$, i.e. we could
identify $\omega_{\hat{a}}= X_{\hat{a}}$. The constraint
(\ref{3.19}) in this case implies that the four-dimensional gauge
group is $U(1)$ so that $A_\mu(x)$ is $U(1)$ valued. The
constraint (\ref{eq7}) leads again to the scalar singlet
$\vphi(x)$.

In general, we start with a $U(1)$ gauge theory on $M^4\times
S^2_F$. We solve the CSDR constraint (\ref{3.17}) by embedding
$SU(2)$ in $U(N)$. There exist $p_{N}$ embeddings, where $p_N$ is
the number of ways one can partition the integer $N$ into a set of
non-increasing positive integers \cite{Mad}. Then the constraint
(\ref{3.19}) gives the surviving four-dimensional gauge group. The
constraint (\ref{eq7}) gives the surviving four-dimensional
scalars and eq. (\ref{soleasy}) is always a solution but in
general not the only one. By setting $\phi_\a=\omega_\a$ we obtain
always a minimum of the potential. This minimum is given by the
chosen embedding of $SU(2)$ in $U(N)$.

In the $G=U(P)$ case,
$\omega_{\hat{a}}=\omega_{\hat{a}}(X^\b)=\omega_\a^{h,\al}T^h{\cal
T}^\al$ is an $N_{}P\times N_{}P$ hermitian matrix and in order
to solve the constraint (\ref{3.17}) we have to embed
${\rm{Lie}}(SU(2))$ in  ${\rm{Lie}}(U(N_{}P))$. All the results
of the $G=U(1)$ case holds also here, we just have to replace
$N$ with $N_{}P$.

One proceeds in a similar way for more
general fuzzy coset $(S/R)_F$ (e.g. fuzzy $CP^M=SU(M+1)/U(M)$)
described by $N \times N$ matrices. The results are again similar,
in particular one starts with a gauge group $G=U(P)$ on $M^4\times
(S/R)_F$, and then the CSDR constraints imply that the
four-dimensional gauge group $K$ is the centralizer of the image
$S_{U(NP)}$  of $S$ in $U(NP)$, $K=C_{U(NP)}(S_{U(NP)})$.

\section{Discussion and Conclusions}
The Fuzzy-CSDR has different features from the ordinary CSDR
leading  therefore to new four-dimensional particle models.  Here
we have stated the rules for the construction of such models; it
may well be that Fuzzy-CSDR provides more realistic
four-dimensional theories. Having in mind the construction of
realistic models one can also combine the fuzzy and the ordinary
CSDR scheme, for example considering $M^4\times S'/{R'}\times
(S/R)_F$.

A major difference between fuzzy and ordinary CSDR is that in
Fuzzy-CSDR the spontaneous symmetry breaking mechanism takes
already place by solving the Fuzzy-CSDR contraints. The four
dimensional Higgs potential has the typical ``mexican hat'' shape,
but it appears already spontaneously broken. Therefore  in four
dimensions appears only the physical Higgs field that survives
after a spontaneous symmetry breaking. Correspondingly in the
Yukawa sector of the theory \cite{us} we obtain the results of the
spontaneous symmetry breaking, i.e. massive fermions and Yukawa
interactions among fermions and the physical Higgs field. We see
that if one would like to describe the spontaneous symmetry
breaking of the SM in the present framework, then one would be
naturally led to large extra dimensions.

A fundamental difference between the ordinary CSDR and its fuzzy
version is the fact that a non-abelian gauge group $G$ is not
really required in high dimensions. Indeed  the presence of a
$U(1)$ in the higher-dimensional theory is enough to obtain
non-abelian gauge theories in four dimensions.
\sk

{\bf Acknowledgment} G.~Z. would like to thank the organizers for
the warm hospitality. P.~M. and G.~Z. acknowledge partial support
by EU under the RTN contract HPRN-CT-2000-00148, the Greek-German
Bilateral Programme IKYDA-2001 and by  the NTUA programme for
fundamental research ``THALES".

\end{document}